\documentclass[showpacs,preprintnumbers,amsmath,amssymb,11pt]{article}
\usepackage{graphicx,amsfonts,amssymb,amsthm,amsmath,graphics,epsfig,pstricks,fancyhdr,fancybox}
\usepackage{dcolumn}
\usepackage{bm}

 \newcommand{\eqd}{\stackrel{d}{=}}
 \newcommand{\rv}{random variable}
 \newcommand{\cdf}{cumulative distribution function}
 \newcommand{\pdf}{density function}
 \newcommand{\chf}{characteristic function}
 \newcommand{\lch}{logarithmic characteristic}
 \newcommand{\iid}{independent and identically distributed}
 \newcommand{\id}{identically distributed}
 \newcommand{\ID}{infinitely divisible}
 \newcommand{\SD}{selfdecomposable}
 \newcommand{\sde}{stochastic differential equation}
 \newcommand{\SST}{strictly stable}
 \newcommand{\ST}{stable}
 \newcommand{\ou}{Ornstein--Uhlenbeck}
 \newcommand{\aST}{$\alpha$--stable}
 \newcommand{\LID}{\mathcal{F}_{ID}}
 \newcommand{\LSD}{\mathcal{F}_{SD}}
 \newcommand{\LST}{\mathcal{F}_{St}}
 \newcommand{\uan}{uniformly, asymptotically negligible}
 \newcommand{\pr}[1]{\mathbf{P}\left\{{#1}\right\}}
 \newcommand{\sign}[1]{\,\mathrm{sign}\left({#1}\right)}
 \newcommand{\ave}[1]{\mathbf{E}\left[{#1}\right]}
 \newcommand{\Pas}{\mathbf{P}\hbox{-a.s.}}
 \newcommand{\CLP}{\emph{CLP}}
 \newcommand{\ton}{\stackrel{n}{\longrightarrow}}
 \newcommand{\refeq}[1]{~(\ref{#1})}
 
\begin{document}

\title{\huge \textbf{Selfdecomposability and selfsimilarity:\\
a concise primer}}
\author{\textsc{Nicola Cufaro Petroni}\\
Dipartimento di Matematica and \textsl{TIRES}, Bari University\\
\textsl{INFN} Sezione di Bari\\
via E. Orabona 4, 70125 Bari, Italy\\
email: cufaro@ba.infn.it}
\date{}
 \maketitle
\begin{abstract}
\noindent We summarize the relations among three classes of laws:
infinitely divisible, selfdecomposable and stable. First we look at
them as the solutions of the Central Limit Problem; then their role
is scrutinized in relation to the L\'evy and the additive processes
with an emphasis on stationarity and selfsimilarity. Finally we
analyze the Ornstein--Uhlenbeck processes driven by L\'evy noises
and their selfdecomposable stationary distributions, and we end with
a few particular examples.
\end{abstract}




{\small \tableofcontents}

\section{Notations and preliminary remarks}\label{introduction}

Selfsimilarity is a very popular research topics since a few years,
and it has been approached from many different standpoints producing
an unavoidable level of confusion~\cite{mccauley}. The present paper
is devoted to a short summary of the properties of some important
and well known families of laws: infinitely divisible,
selfdecomposable and stable (for details see for
example~\cite{loeve1,loeve2,gnedenko,sato}). First of all we will
recall their role in the formulation and in the solutions of the
central limit problem: as we will see in the next section this
amounts to a quest for all the limit laws of sums of independent \rv
s. Our families of distributions will then be analyzed by means of
both their possible decompositions in other laws, and the explicit
form of their \chf s: the celebrated L\'evy--Khintchin formula. In
particular it will be discussed the intermediate role played by the
selfdecomposable laws between the more popular stable, and
infinitely divisible distributions. We will then explore these laws
in connection with the additive and the L\'evy processes, looking
for their importance with respect to both the properties of
stationarity and selfsimilarity. In particular it will be recalled
how from selfdecomposable distributions it is always possible to
define both stationary and selfsimilar additive processes which --
with the exception of important particular cases -- will in general
be different. A few remarks are also added to show the differences
between this selfsimilarity and that of the well known fractional
Brownian motion. We will also analyze the Ornstein--Uhlenbeck
processes driven by these L\'evy noises, and their stationary
distributions which are always selfdecomposable. We will finally
elaborate a few examples to illustrate these results, and to compare
the behavior of our processes.

In what follows we will adopt the following notations (for details see for example~\cite{loeve1,shiryayev}): $X,
Y, \ldots$ will denote the random variables, namely the measurable functions $X(\omega), Y(\omega), \ldots$
defined on a probability space $(\Omega,\mathcal{F},\mathbf{P})$ where $\Omega$ is a sample space, $\mathcal{F}$
is a $\sigma$--algebra of events and $\mathbf{P}$ a probability measure. We will then respectively write $F(x),
G(y), \ldots$ for their \cdf s
\begin{equation*}
    F(x)=\mathbf{P}\{X\leq x\}\,,\quad G(y)=\mathbf{P}\{Y\leq y\}\,,\quad\ldots
\end{equation*}
and $\varphi(u), \chi(v), \ldots$ for their \chf s
\begin{equation*}
    \varphi(u)=\mathbf{E}\left(e^{iuX}\right)\,,\quad\chi(v)=\mathbf{E}\left(e^{ivY}\right)\,,\quad\ldots
\end{equation*}
where the symbol $\mathbf{E}$ denotes the expectation value of a \rv\ according to the probability $\mathbf{P}$,
namely for example
\begin{equation*}
    \mathbf{E}(X)=\int_{\Omega}X(\omega)\,d\mathbf{P}=\int_{-\infty}^{+\infty}x\,F(dx)\,.
\end{equation*}
When they exist, $f(x), g(y), \ldots$ will be the probability density functions, and in that event we will have
\begin{equation*}
    F(x)=\int_{-\infty}^xf(z)\,dz\,,\qquad f(x)=F'(x)\,,\qquad\mathbf{E}(X)=\int_{-\infty}^{+\infty}xf(x)\,dx\,.
\end{equation*}
A law will be indifferently specified either by its \cdf\ (or \pdf), or by its \chf. To say that a \rv\ $X$ is
distributed according to a given law we will also write either $X\sim F(x)$ or $X\sim\varphi(u)$; if a family of
laws is denoted by a specific symbol, say $\mathcal{L}$, then we also write $X\sim\mathcal{L}$ to say that $X$ is
distributed according to one of these laws. When two \rv s $X$ and $Y$ are identically distributed we will adopt
the notation
\begin{equation*}
    X\eqd Y\,.
\end{equation*}
If a property is true with probability 1 we will also say that it is true $\mathbf{P}$--\emph{almost surely} and
we will adopt the notation $\Pas$\ A \emph{type of laws} (see~\cite{loeve1} Section 14) is a family of laws that
only differ by a centering and a rescaling: in other words, if $\varphi(u)$ is the \chf\ of a law, all the laws
of the same type have \chf s $e^{ibu}\varphi(au)$ with a centering parameter $b\in\mathbb{R}$, and a scaling
parameter $a>0$ (we exclude here the sign inversions). In terms of \rv s this means that the laws of $X$ and
$aX+b$ (for $a>0$, and $b\in\mathbb{R}$) always are of the same type, and on the other hand that $X$ and $Y$
belong to the same type if it is possible to find $a>0$, and $b\in\mathbb{R}$ such that $Y\eqd aX+b$. We will
also say that a \rv\ $X\sim\varphi(u)$ and its law are \emph{composed} of $X_1\sim\varphi_1(u)$ and
$X_2\sim\varphi_2(u)$ when $X_1$ and $X_2$ are independent and $X\eqd X_1+X_2$, or equivalently when
$\varphi(u)=\varphi_1(u)\varphi_2(u)$; then $X_1$ and $X_2$ are also called \emph{components} of $X$. Of course,
in terms of distributions, a composition amounts to a \emph{convolution}; then, for example, if
$\mathcal{N}(a^2,b)$ denotes a normal law with expectation $b$ and variance $a^2$, the composition of two normal
laws will also be indicated as $\mathcal{N}(a_1^2,b_1)*\mathcal{N}(a_2^2,b_2)$. For the stochastic processes
$X(t), Y(t),\ldots$ we will say that $X(t)$ and $Y(t)$ are \emph{identical in law} when the systems of their
finite--dimensional distributions are identical, and in this case we will write
\begin{equation*}
    X(t)\eqd Y(t)\,.
\end{equation*}
A process $X(t)$ is said to be \emph{stochastically continuous} if
for every $\epsilon>0$
\begin{equation*}
    \lim_{\Delta t\rightarrow0}\pr{|X(t+\Delta
    t)-X(t)|>\epsilon}=0,\qquad\forall\,t\geq0.
\end{equation*}
where $\pr{\ldots}$ denotes the probability for the increment $|X(t+\Delta t)-X(t)|$ of being larger that
$\epsilon>0$. In this paper all our \rv s and processes will be one dimensional.

In this exposition we do not pretend neither rigor, nor
completeness: we just list the results and the properties that are
important to compare, and we refer to the existing literature for
proofs and details, and for a few hints about possible recent
applications. In fact the aim of this paper is just to draw an
outline showing -- without embarrassing the reader with excessive
technical details -- the deep, but otherwise simple ideas which are
behind the properties of our families of laws and the simplest
procedures to build from them the most important classes of
processes. We hope that this, with the aid of some telling examples,
will be helpful to approach this field of research by clarifying the
roles, the differences and the subtleties of these laws and
processes.

\section{The Central Limit Problem}

\subsection{The classical limit theorems}\label{classicalth}

It is well known that there are three kinds of \emph{classical limit
theorems} (all along this paper wherever we speak of convergence it
is understood that we speak of \emph{convergence in law}) which are
characterized by the form of the respective limit laws:
\begin{itemize}
    \item the \emph{Law of Large Numbers} with limit laws of
    the degenerate type $\delta_b$ with \chf\
    \begin{equation*}
        \varphi(u)=e^{ibu};
    \end{equation*}
    \item the \emph{Normal Central Limit Theorem} whose limit laws
    are of the gaussian type $\mathcal{N}(a^2,b)$ with
    \begin{equation*}
        \varphi(u)=e^{ibu}e^{-a^2u^2/2};
    \end{equation*}
    \item and the \emph{Poisson Theorem} whose limit laws are of
    the Poisson types $\mathcal{P}(\lambda;a,b)$ with
    \begin{equation}\label{poisschf}
        \varphi(u)=e^{ibu}e^{\lambda(e^{iau}-1)}.
    \end{equation}
\end{itemize}
The exact statements of these theorems in their traditional
formulations are reprinted in every handbook of probability (see for
example~\cite{shiryayev} p.\ 323 and following), and we will not
reproduce them once more. Remark however that in our list, while
speaking of \textit{the type} for the degenerate and the gaussian
laws, we also referred to \textit{the types} for the Poisson laws.
In fact all the normal laws $\mathcal{N}(a^2,b)$ with expectation
$b\in\mathbb{R}$ and variance $a^2>0$ constitute a unique type, and
the same is true for the family of all the degenerate laws. On the
other hand the standard Poisson laws
$\mathcal{P}(\lambda)=\mathcal{P}(\lambda;1,0)$ with different
parameters $\lambda$ belong to different types: indeed we can not
recover a law $\mathcal{P}(\lambda)$ from another
$\mathcal{P}(\lambda')$ (with $\lambda\neq\lambda'$) just by means
of a centering and a rescaling. In other words we could say that
every Poisson law $\mathcal{P}(\lambda)$ with a given $\lambda$
generates -- by centering and rescaling -- a distinct type
$\mathcal{P}(\lambda;a,b)$ with \chf s\refeq{poisschf}.

The particularities of the Poisson laws with respect to the other
two families of limit laws are also apparent from their properties
of \emph{composition and decomposition}. For a family of laws (not
necessarily a type) to be closed under composition means that the
composition (convolution) of two laws of that family still belongs
to the same family. On the other hand closure under decomposition
means that if a law of the family is decomposed in two laws, these
two components necessarily belong to the same family. The types of
limit laws appearing in the classical limit theorems show an
important form of closure under composition and decomposition
summarized in the following result (see~\cite{loeve1} p.\ 283):
\emph{the degenerate and normal types are closed under compositions
and under decompositions; the same is true for every family of
Poisson laws $\mathcal{P}(\lambda;a,b)$ with the same $a$.} The
closure under composition is in fact an elementary property; not so
for the closure under decomposition: the proofs for the normal and
the Poisson case were given in 1935-37 by H.\ Cram\'er and D.A.\
Raikov respectively. The normal and Poisson composition and
decomposition properties can then be stated by saying that
\begin{eqnarray*}
  \mathcal{N}(a_1^2,b_1)*\mathcal{N}(a_2^2,b_2) &=& \mathcal{N}(a_1^2+a_2^2,b_1+b_2) \\
  \mathcal{P}(\lambda_1;a,b_1)*\mathcal{P}(\lambda_2;a,b_2) &=& \mathcal{P}(\lambda_1+\lambda_2;a,b_1+b_2)
\end{eqnarray*}
Hence it is also apparent that a type of Poisson laws never is
closed under composition and decomposition: we are always obliged to
switch from a Poisson type to another while composing and
decomposing them.

It is important to remark, however, that the classical limit
theorems are not the embodiment of these composition and
decomposition properties only, but they are more far--reaching and
profound statements. In fact not only these theorems deal with
limits of sums of independent \rv s of a type which in general is
different from that of the eventual limit laws -- the Poisson law is
the limit of sums of Bernoulli 0--1 \rv s, while the normal and
degenerate laws are limits of sums of still more general \rv s --
but also the distribution of the sum of the $n$ \rv s does not
coincide with the limit law at every step $n$ of the limiting
process, as happens instead in a simple decomposition.

\subsection{Formulations of the Central Limit Problem}

To formulate the Central Limit Problem (\CLP) let us look at it
first of all in terms of sequences of \rv s: usually we take a
sequence $X_k$ of \rv s, and then the sequence of their sums
$S_n=X_1+\ldots+X_n$. In this case when we go from $S_n$ to
$S_{n+1}$ we just add another \rv\ $X_n$ without changing the
previous sum $S_n$. However this is not the more general way to
produce sequences of sums of \rv s. Consider indeed a triangular
array of \rv s $X_k^{(n)}$
\[
\begin{array}{l}
     X_1^{(1)}\\
     X_1^{(2)}\,,\;\;X_2^{(2)}\\
     \;\;\vdots\qquad\qquad\quad \ddots\\
     X_1^{(n)}\,,\;\;X_2^{(n)}\;,\;\ldots\;,\;\;X_n^{(n)}\\
     \;\;\vdots\qquad\qquad\qquad\qquad\qquad\ddots\\
\end{array}
\]
with $n=1,2,\ldots$ and $k=1,2,\ldots,n$, and suppose that
\begin{enumerate}
    \item in every row $n\in\mathbb{N}$ the \rv s
    $X_1^{(n)},\ldots,X_{n}^{(n)}$ are independent,
    \item the $X_k^{(n)}$ are \emph{uniformly, asymptotically negligible}, namely that
    \begin{equation*}
        \max_k\,\pr{|X_k^{(n)}|\geq\epsilon}\ton0\,,\qquad\forall\,\epsilon>0.
    \end{equation*}
\end{enumerate}
Define now the \emph{consecutive sums}
\begin{equation}\label{consecsum}
    S_n=\sum_{k=1}^{n}X_k^{(n)}\,.
\end{equation}
Then the central limit problem \emph{for consecutive sums of
independent \rv s} (\CLP$_1$) reads: \emph{find the family of all
the limit laws of the consecutive sums}\refeq{consecsum} \emph{and
the corresponding convergence conditions} (see~\cite{loeve1} p.\
301-2). Three remarks are in order here:
\begin{itemize}
    \item for a given $n\in\mathbb{N}$ the \rv s $X_1^{(n)},\ldots,X_{n}^{(n)}$ are
    independent, but in general they are neither \id, nor of the same
    type;
    \item going from the row $n$ to the row $n+1$ the \rv s and their laws
    change: in general $X_k^{(n)}$ and $X_k^{(n+1)}$ are neither \id, nor of the same
    type; as a consequence going from $S_n$ to $S_{n+1}$ we not only add the $n+1$-th \rv, but we also are obliged
    to adjourn the law of $S_n$;
    \item the \uan\ condition is an important technical requirement added to
    avoid trivial answers to the \CLP$_1$ (see for example~\cite{loeve1} p.
    302); in fact without this condition it is easy to show
    that any law $\varphi$ would be limit law of the consecutive
    sums\refeq{consecsum}: it would be enough for every $n$ to take $X_1^{(n)}\sim\varphi$, and
    $X_k^{(n)}=0\,\;\Pas$ for $k>1$. The \uan\ condition will tacitly be assumed all along this paper.
\end{itemize}
Important particular cases of the \CLP$_1$ are then selected when we
specialize the sequence $X_k^{(n)}$ in the following way
(see~\cite{loeve1} p.\ 331): let us suppose that there is a sequence
$X_k$, $k=1,2,\ldots$, of independent (but in general not \id) \rv
s, and two sequences of numbers $a_n>0$ and $b_n\in\mathbb{R}$,
$n=1,2,\ldots$, such that for every $k$ and $n$
    \begin{equation}\label{specialseq}
        X_k^{(n)}=\frac{1}{a_n}\left(X_k-\frac{b_n}{n}\right).
    \end{equation}
It is apparent that now going from $n$ to $n'$ we just get a
centering and a rescaling of every $X_k$, so that $X_k^{(n)}$ and
$X_k^{(n')}$ always are of the same type. Then the consecutive sums
take the form of \emph{normed sums} (namely centered and rescaled
sums) of independent \rv s
\begin{equation}\label{normsum}
    S_n=\sum_{k=1}^nX_k^{(n)}=\sum_{k=1}^n\frac{1}{a_n}\left(X_k-\frac{b_n}{n}\right)=
    \frac{1}{a_n}\left(\sum_{k=1}^nX_k-b_n\right)=\frac{\widetilde{S}_n-b_n}{a_n}
\end{equation}
where we adopt the notation
\begin{equation*}
    \widetilde{S}_n=\sum_{k=1}^nX_k.
\end{equation*}
Then the central limit problem\ \emph{for normed sums of independent
\rv s} (\CLP$_2$) reads: \emph{find the family of all the limit laws
of the normed sums}\refeq{normsum} \emph{and the corresponding
convergence conditions}. Finally there is a still more specialized
formulation of the central limit problem\ when we add the hypothesis
that the \rv s $X_k$ are not only independent, but also identically
distributed (see~\cite{loeve1} p.\ 338): in this case we speak of a
central limit problem\ \emph{for normed sums of \iid\ \rv s}
(\CLP$_3$).

\subsection{Solutions of the Central Limit Problem}

The answers to the different formulations of the central limit
problem\ need the definition of several important families of laws
that are much more general than the Gaussian type, and that can be
defined by means of the properties of their \chf s. Suppose that
$\varphi(u)$ is the \chf\ of a law: we will say that this law is
\emph{infinitely divisible} (see~\cite{loeve1} p.\ 308) if for every
$n\in\mathbb{N}$ we can always find another \chf\ $\varphi_n(u)$
such that
\begin{equation*}
    \varphi(u)=[\varphi_n(u)]^n.
\end{equation*}
Apparently the name comes from the fact that our law can always be
decomposed in an arbitrary number of identical laws; remark however
that for different values of $n$ we get in general laws
$\varphi_n(u)$ of different types. In terms of \rv s if
$X\sim\varphi(u)$ is \ID, then for every $n\in\mathbb{N}$ we can
find $n$ \iid\ \rv s $X_1^{(n)},\ldots,X_n^{(n)}$ all distributed as
$\varphi_n(u)$ and such that
\begin{equation*}
    X\eqd\sum_{k=1}^nX_k^{(n)},
\end{equation*}
namely $X$ is always decomposable (in distribution) in the sum of an
arbitrary, finite number of \iid\ \rv s. Let us call $\LID$ the
family of all the \ID\ laws. Many important distributions are \ID:
degenerate, Gaussian, Poisson, compound Poisson, geometric, Student,
Gamma, exponential and Laplace are \ID. On the other hand the
uniform, Beta and binomial laws are not \ID: in fact no distribution
(other than the degenerate) with bounded support can be \ID\
(see~\cite{sato} p.\ 31). Remark that if $\varphi(u)$ is an \ID\
\chf, then also $\varphi^\lambda(u)$ is an \ID\ \chf\ for every
$\lambda>0$ (see~\cite{sato} p.\ 35): we will see that this is
instrumental to connect the \ID\ laws to the L\'evy processes.

A second important family of laws selected by their decomposition
properties is that of the \emph{selfdecomposable} laws
(see~\cite{loeve1} p.\ 334): a law $\varphi(u)$ is \SD\ when for
every $a\in(0,1)$ we can always find another \chf\ $\varphi_a(u)$
such that
\begin{equation}\label{sdlaw}
    \varphi(u)=\varphi(au)\varphi_a(u).
\end{equation}
In terms of \rv s this means that if $X\sim\varphi(u)$ is \SD, then
for every $a\in(0,1)$ we can always find two independent \rv s,
$X'\eqd X$ and $Y_a\sim\varphi_a(u)$, such that
\begin{equation*}
    X\eqd aX'+Y_a\,.
\end{equation*}
In other words for every $a\in(0,1)$ $X$ can always be decomposed
into two independent \rv s such that one of them is of the same type
of $X$. It can be shown that every \SD\ law, along with all its
components, is also \ID\ (see~\cite{loeve1} p.\ 335), so that if we
call $\LSD$ the family of all the \SD\ laws, then
$\LSD\subseteq\LID$. The Gaussian, Student, Gamma, exponential and
Laplace laws are examples of \SD\ laws (see~\cite{sato} p.\ 98). On
the other hand the Poisson laws are not \SD: they only are \ID.

Finally we will say that a law $\varphi(u)$ is \emph{stable}
(see~\cite{loeve1} p.\ 338) if for every $c_1>0$ and $c_2>0$ we can
find $a>0$ and $b\in\mathbb{R}$ such that
\begin{equation*}
    e^{ibu}\varphi(au)=\varphi(c_1u)\varphi(c_2u).
\end{equation*}
This means now that if $X\sim\varphi(u)$ is \ST, then  for every
$c_1>0$ and $c_2>0$ we can find two independent \rv s $X_1\eqd X$
and $X_2\eqd X$, and two number $a>0$ and $b$ such that
\begin{equation*}
    aX+b\eqd c_1X_1+c_2X_2.
\end{equation*}
In other words we can always decompose a \ST\ law in other laws
which are of the same type as the initial one. Our definition can
also be reformulated in a slightly different way (see~\cite{sato}
p.\ 69): for every $c>0$ we can always find $a>0$ and
$b\in\mathbb{R}$ such that
\begin{equation*}
    e^{ibu}\varphi(au)=[\varphi(u)]^c,
\end{equation*}
and this apparently means that for every $c>0$ the law
$[\varphi(u)]^c$ is of the same type as $\varphi(u)$. A law is also
said \emph{strictly stable} if for every $c>0$ always exists $a>0$
such that
\begin{equation}\label{stable}
    \varphi(au)=[\varphi(u)]^c.
\end{equation}
All the \ST\ laws are \SD, and hence if $\LST$ is the family of all
the \ST\ laws we will have $\LST\subseteq\LSD$. In fact our
classification of laws in only three families (\ID, \SD\ and \ST) is
an oversimplification of a much richer structure  explored for
example in~\cite{sato}, Chapter 3. Among the classical laws only the
Gaussian and the Cauchy laws are \ST.

Remark that if a law $\varphi$ belongs to one of our families, then
also all its type belongs to the same family. Hence it would be more
suitable to say that $\LID$, $\LSD$ and $\LST$ are families of types
of laws; however in the following, for the sake of simplicity, this
will be understood without saying. The relevance of our three
families of laws lies in the fact that they represent the answers to
the three formulations of the central limit problem\ discussed in
the previous section. In fact it can be shown that $\LID\,,\, \LSD$
\emph{and} $\LST$ \emph{exactly coincide with the families of the
limit laws sought for respectively in} \CLP$_1$, \CLP$_2$\
\emph{and} \CLP$_3$ (see for example~\cite{loeve1} p.\ 321, p.\ 335
and p.\ 339 for the three statements). To summarize these results --
a few examples will be shown in the Section~\ref{exampleCLT} -- we
can then say that:
\begin{itemize}
    \item the laws of the normed sums\refeq{normsum} of sequences $X_k$ of \iid\
    \rv s converge toward \ST\ laws; in particular: when the
    $X_k$ have finite variance their normed sums converge -- according to the classical theorem -- to normal
    laws, while the non Gaussian, stable distributions are limit laws only
    for sums of \rv s with infinite variance; a well known
    example of the non Gaussian limit laws is the Cauchy
    distribution;
    \item the laws of the normed sums\refeq{normsum} of sequences $X_k$ of
    independent, but not necessarily \id, \rv s converge toward \SD\
    laws; the special case of the stable laws is recovered when the
    $X_k$ are also \id; in other words when a \SD\ law is not
    \ST\ it can not be the limit law of normed sums of \iid\ \rv s;
    \item finally the laws of the consecutive sums\refeq{consecsum}
    of triangular arrays $X_k^{(n)}$ converge to \ID\ laws: a
    classical example of non \SD\ convergence is represented by the Poisson
    Limit Theorem recalled in the Section~\ref{classicalth}; of course the \SD\ case is obtained when the
    triangular array has the form\refeq{specialseq} and the
    consecutive sums become normed sums\refeq{normsum}; however \ID\
    laws which are not \SD\ (as the Poisson law) can not be limit laws of normed
    sums of independent \rv s.
\end{itemize}
Nothing forbids, of course, that in this scheme a Gaussian law be
also the limit law for consecutive sums of triangular arrays that do
not reduces to normed sums. To complete the picture we will hence
just recall here that there are also necessary and sufficient
conditions for the convergence to normal laws of triangular arrays
of independent, but not necessarily \id, \rv s with finite variances
(see~\cite{shiryayev} p.\ 326).

Since the \ST\ distributions are limit laws of normed
sums\refeq{normsum} of \iid\ \rv s we can also introduce the notion
of \emph{domain of attraction of a} \ST\ \emph{law} $\varphi_S\,$:
we will say that a law $\varphi$ belongs to the domain of attraction
of $\varphi_S$ when we can find two sequences of numbers, $a_n>0$
and $b_n$, such that the normed sums\refeq{normsum} of a sequence
$X_k$ of \rv s \emph{all distributed as} $\varphi$, converge to
$\varphi_S$. Remark that this definition can not be immediately
extended to the non \ST\ distributions which are not limit laws of
normed sums of \iid\ \rv s, so that we can not speak of a
\emph{unique} distribution $\varphi$ being attracted by the limit
law.

Every law with finite variance belongs to the domain of attraction
of the normal law (see~\cite{loeve1} p.\ 363). It is also important
to stress here that, while all \ST\ laws are attracted by themselves
(see~\cite{loeve1} p.\ 363), a non \ST, \ID\ law $\varphi$ -- which
always is in itself the limit law of a suitable consecutive
sum\refeq{consecsum} of some triangular array of \rv s -- also
belongs to the domain of attraction of some stable law $\varphi_S$:
normed sums\refeq{normsum} of \rv s \emph{all distributed as}
$\varphi$ will converge toward some stable law $\varphi_S$. For
instance the Poisson law -- which is an \ID\ limit law, as the
Poisson Theorem shows -- apparently also is in the domain of
attraction of the normal law since it has a finite variance: a
normed sum of \rv s all distributed according to the same Poisson
law will converge to the Gauss law. Finally we recall, without going
into more detail, that for a given law $\varphi$ it is always
possible to find if it belongs to some domain of attraction, and
then it is also possible to find both the \ST\ limit law
$\varphi_S$, and the admissible numerical sequences $a_n>0$ and
$b_n$ entering in the normed sums\refeq{normsum} converging to
$\varphi_S$ (see~\cite{loeve1} p.\ 364).

\subsection{The L\'evy--Khintchin formula}

It is not easy to find out if a given law $\varphi(u)$ belongs to
one of the families defined in the previous section just by looking
at the definitions introduced up to now. It is important then to
recall the explicit form of the \chf s of our families of laws given
by the \emph{L\'evy--Khintchin formula}. It can be proved
(see~\cite{loeve1} p.\ 343) indeed that the \emph{logarithmic
characteristic} $\psi(u)=\log\varphi(u)$ of an \ID\ law is uniquely
associated, through the formula
\begin{equation}\label{LKformula}
  \psi(u)=iu\gamma-\frac{\beta^2}{2}\,u^2
           +\lim_{\delta\rightarrow0}\int_{|x|>\delta}\left(e^{iux}-1-\frac{iux}{1+x^2}\right)\,dL(x)
\end{equation}
to a \emph{generating triplet} $\left(\beta^2,L,\gamma\right)$ where
$\gamma,\,\beta\in\mathbb{R}$, and the \emph{L\'evy function} $L(x)$
is defined on $\mathbb{R}\backslash\{0\}$, is non decreasing on
$(-\infty,0)$ and $(0,+\infty)$, with $L(\pm\infty)=0$ and
\begin{equation*}
  \lim_{\delta\rightarrow0}\int_{\delta<|y|<x}y^2\,dL(y)<+\infty
\end{equation*}
for $x>0$ finite. In other words a law will be \ID\ if and only if
its \lch\ satisfies the relation\refeq{LKformula} for a suitable
choice of $\left(\beta^2,L,\gamma\right)$. The L\'evy function
$L(x)$ also defines the \emph{L\'evy measure} (for details
see~\cite{sato} Section 8)
\begin{equation*}
    \nu(B)=\int_BdL(x)
\end{equation*}
for every measurable set $B$ of $\mathbb{R}$ and
$\nu\left(\{0\}\right)=0$, so that often we will refer to
$\left(\beta^2,\nu,\gamma\right)$ as the generating triplet. When
the L\'evy measure $\nu$ is absolutely continuous we will also
denote by $W(x)=L'(x)$ its density. Remark that the
L\'evy--Khintchin formula\refeq{LKformula} can be given in a variety
of equivalent versions (see~\cite{sato} p.\ 37) by suitably choosing
the integrand functions, but we will not go into such details.

For a given law even the verification of the
formula\refeq{LKformula} is not in general an easy task. In the case
of \ST\ laws, however, the L\'evy--Khintchin formula is considerably
simpler since it no longer explicitely involves integrals on the
L\'evy measure. A \ST\ law is characterized by a parameter
$0<\alpha\leq2$ (see~\cite{sato} p.\ 76) and it is then also said
\aST\ ($\alpha$-stable): the case $\alpha=2$ corresponds to the
Gaussian laws, and to a vanishing L\'evy measure. In the non
Gaussian \aST\ cases ($0<\alpha<2$) on the other hand the L\'evy
measure is not zero, it is absolutely continuous and we have
(see~\cite{sato} p.\ 80)
\begin{equation*}
   dL(x)=W(x)\,dx=\left\{
           \begin{array}{ll}
             A\,x^{-1-\alpha}dx, & \hbox{for $x>0$,} \\
             B|x|^{-1-\alpha}dx, & \hbox{for $x<0$,}
           \end{array}
         \right.
\end{equation*}
with $A\geq0,\;B\geq0$ and $A+B>0$. Finally in both cases --
Gaussian and non Gaussian -- the \lch\ must satisfy the following
relation (see~\cite{sato} p.\ 86)
\begin{equation*}
  \psi(u)=
  \begin{cases}
    ia u-b|u|^\alpha\left(1-i\sign{u}c\tan\frac{\pi}{2}\alpha\right) & \text{if $\alpha\neq1$}, \\
    ia u-b|u|\left(1+i\sign{u}\frac{2}{\pi}\,c\log|u|\right) & \text{if
    $\alpha=1$},
  \end{cases}
\end{equation*}
where $\alpha\in(0,2]$, $a\in\mathbb{R}$, $b>0$ and $|c|\leq1$. The
Gaussian case simply corresponds to $\alpha=2$. When the law is also
symmetric the \chf\ is real and the formula reduces itself to the
quite elementary expression
\begin{equation}\label{symmstable}
  \varphi(u)=e^{-b|u|^\alpha}\,,\qquad 0<\alpha\leq2\,.
\end{equation}

The form of the L\'evy--Khintchin formula, or equivalently of the triplet $\left(\beta^2,L,\gamma\right)$, of the
\SD\ laws, on the other hand, is not so simple. They in fact play in some sense a sort of intermediate role
between the generality of the \ID\ laws and the special properties of the \ST\ laws. It can be proved indeed
(see~\cite{sato} p.\ 95) that a law is \SD\ if and only if its L\'evy measure is absolutely continuous and its
density is
\begin{equation}\label{sdlevymeas}
    W(x)=L'(x)=\frac{k(x)}{|x|}
\end{equation}
where the function $k(x)$ is non negative, is increasing on $(-\infty,0)$ and is decreasing on $(0,+\infty)$. By
the way this also show why a Poisson law (whose L\'evy measure is not absolutely continuous) can not be \SD. As a
consequence the L\'evy--Khintchin formula~\refeq{LKformula} of the \SD\ laws is more specialized than that of the
general \ID\ laws, but it still contains a non elementary integral part.

\section{L\'evy processes and additive processes}

An \emph{additive process} $X(t)$ is a stochastically continuous
process with independent increments and $X(0)=0\,,\;\Pas$ (namely
the probability of not being zero vanishes); on the other hand a
\emph{L\'evy process} is an additive process with the further
requirement that the increments must be stationary (for further
details see~\cite{sato} Section 1). The \emph{stationarity of the
increments} means that the law of $X(s+t)-X(s)$ does not depend on
$s$. Processes with independent increments are also Markov
processes, and hence the entire family of their joint laws at an
arbitrary, finite number of times can be deduced just from the one--
and the two--times distributions. In other words it is enough to
know the laws of the increments to have the complete law of the
process. This of course is a very good reason to be interested in
Markov, and in particular in additive processes, but it must be
recalled here that there are also non Markovian processes which can
still be defined by means of very simple tools. An important example
that will be briefly mentioned later is the \emph{fractional
Brownian motion} which takes advantage of being a Gaussian process
to make up for its lack of Markovianity. It is also important to
recall here that -- with the exception of Gaussian processes -- the
additive processes trajectories can make jumps. This does not
contradict their stochastic continuity because the jumping times are
random, and hence, for every $t$, the probability of a jump
occurring exactly at $t$ is zero.

\subsection{Stationarity and \ID\ laws}

In the following the law of the process increment $X(t)-X(s)$ will
be given by means of its \chf\ $\phi_{s,t}(u)$, so that the
stationarity of the L\'evy processes simply entails that
$\phi_{s,t}(u)$ only depends on the difference $\tau=t-s$: in this
case we will use the shorthand, one--time notation $\phi_\tau(u)$.
As for every Markov process the laws of the increments of an
additive process must satisfy the Chapman--Kolmogorov equations
which for the \chf s are
\begin{equation}\label{chk}
    \phi_{r,t}(u)=\phi_{r,s}(u)\phi_{s,t}(u)\,,\qquad0\leq r<s<t\,;
\end{equation}
for L\'evy (stationary) processes these equations take the form
\begin{equation}\label{chkstat}
    \phi_{\sigma+\tau}(u)=\phi_\sigma(u)\phi_\tau(u)\,,\qquad\sigma,\tau>0\,.
\end{equation}

There is now a very simple and intuitive procedure to build a L\'evy
process: take the \chf\ $\varphi(u)$ of a law and define
\begin{equation}\label{stchf}
    \phi_t(u)=\left[\varphi(u)\right]^{t/T}.
\end{equation}
It is immediate to see that $\phi_t(u)$ satisfies\refeq{chkstat}, so
that it can surely be taken as the \chf\ of the stationary
increments of a L\'evy process. Here $T$ plays the role of a
dimensional time constant (a \emph{time scale}) introduced to have a
dimensionless exponent. To have a consistent procedure, however, we
must be sure that when $\varphi(u)$ is a \chf, also $\phi_t(u)$
in\refeq{stchf} is a \chf\ for every $t>0$, but unfortunately this
is simply not true for every \chf\ $\varphi(u)$. We are led hence to
ask for what kind of \chf s\refeq{stchf} is again a \chf. We know,
on the other hand, that if $\varphi$ is an \ID\ \chf, then also
$\varphi^\lambda$ with $\lambda>0$ is a \chf, and an \ID\ one too.
In fact it is possible to show that\refeq{stchf} is a \chf\ if and
only if $\varphi(u)$ is \ID. In other words there is a one-to-one
relation between the class $\LID$ of the \ID\ laws and that of the
L\'evy processes (see~\cite{sato} Section 7). Remark however that in
general for a L\'evy process defined by\refeq{stchf} the \ID\ law of
the increments at a generic time $t$ is neither $\varphi(u)$, nor of
the same type of $\varphi(u)$. Only at $t=T$ the law is necessarily
$\varphi(u)$, while for $t\neq T$ it can be rather different and --
but for few well known cases -- its explicit \cdf\ (or \pdf) could
be quite difficult to find.

On the other hand when $\varphi(u)$ is a \ST\ law it is easy to see
from the very definition\refeq{stable} of stability that at every
time $t$ the law of the increments\refeq{stchf} will always belong
to the same type (this is famously what happens for the Gauss and
Cauchy laws). In this case we speak of a \ST\ process, and its
evolution can be summarized just in the time dependence of the law
parameters which will produce a trajectory inside a unique type. A
different situation arises instead when $\varphi(u)$ only belongs to
a family of \ID\ laws closed under composition and decomposition. As
we have already remarked these families do not in general constitute
a type (as the family of the Poisson laws $\mathcal{P}(\lambda)$):
if however they are closed under composition (as are both the
Poisson and the Compound Poisson processes) the law\refeq{stchf} of
the increment of the L\'evy process stays in the same family of laws
all along an evolution which is described by the time dependence of
the law parameters; this however does not amount to the stability of
the process since our family is not a single type.

Since every L\'evy process is associated to an \ID\ law
$\varphi(u)=e^{\psi(u)}$, and since every \ID\ law is associated to
a generating triplet $\left(\beta^2,\nu,\gamma\right)$ we will also
speak of the \lch\ $\psi(u)$ and of the generating triplet
$\left(\beta^2,\nu,\gamma\right)$ of a L\'evy process. In this case
however the L\'evy measure $\nu$ has also an important probabilistic
meaning w.r.t.\ the L\'evy process (see for example~\cite{cont} pp.\
75-85): \emph{for every Borel set $A$ of $\,\mathbb{R}$, $\nu(A)$
represents the expected number, per unit time, of (non-zero) jumps
with size belonging to $A$}. It can also be proved that for every
compact set A such that $0\notin A$ we have $\nu(A)<+\infty$, namely
the number of jumps per unit time of finite (neither infinite, nor
infinitesimal) size is finite. Remark however that this does not
mean that $\nu$ is a finite measure on $\mathbb{R}$: in fact the
function $L(x)$ associated to $\nu$ can diverge in $x=0$ so that the
process can have an infinite number of infinitesimal jumps in every
compact $[0,T]$. In this case, when $\nu(\mathbb{R})=+\infty$, we
speak of an \emph{infinite activity} process, and the set of the
jump times of every trajectory will be countably infinite and dense
in $[0,+\infty]$. For the sake of simplicity we will not introduce
here the important \emph{L\'evy--It\^o decomposition} of a L\'evy
process into its continuous (Gaussian) and jumping (Poisson) parts:
the readers are referred to~\cite{cont}, Section 3.4 for a synthetic
treatment.

\subsection{Selfsimilarity and \SD\ laws}

A process $X(t)$ (possibly neither additive, nor stationary) is said
to be \emph{selfsimilar} when for every given $a>0$ we can find
$b>0$ such that
\begin{equation*}
    X(at)\eqd bX(t),
\end{equation*}
namely when every change $a$ in the time scale can be compensated
\emph{in distribution} by a corresponding change $b$ in the space
scale. In terms of the increment \chf s this means that for every
$a>0$ we must have a $b>0$ such that
\begin{equation}\label{ssim}
    \phi_{as,at}(u)=\phi_{s,t}(bu).
\end{equation}
In fact it can be proved more about the form of this space--time
compensation: given a
selfsimilar process we can always find $H>0$ such that $b=a^H$
(see~\cite{sato} p.\ 73). This number $H$ is called the
\emph{exponent of the process} or \emph{Hurst index}, and we will
also speak of
$H$\emph{--selfsimilar} processes. For further details about
selfsimilar, additive processes see also~\cite{sato91}.

Since a L\'evy process is completely specified by\refeq{stchf} as
\chf\ of its increments, then in this case the selfsimilarity means
that for every $a>0$ it exists $b>0$ such that
\begin{equation}\label{statssim}
    [\varphi(u)]^{at/T}=[\varphi(bu)]^{t/T}.
\end{equation}
From the definition\refeq{stable} of the \SST\ laws and
from\refeq{statssim} it is easy to understand then that the unique
selfsimilar L\'evy processes must be \SST. For instance in a Wiener
process we have $\varphi(u)=e^{-u^2\sigma^2/2}$, namely
\begin{equation*}
    [\varphi(u)]^{t/T}=e^{-u^2Dt/2},\qquad D=\frac{\sigma^2}{T}\,,
\end{equation*}
and hence
\begin{equation*}
    [\varphi(u)]^{at/T}=e^{-u^2Dat/2}
\end{equation*}
so that $b=\sqrt{a}$ (namely $H=1/2$) is the required compensation.
This means that, insofar as the coefficient $D$ remains the same, we
can change the space and time scales $\sigma$ and $T$ (namely we can
change the units of measure) without changing the Wiener process
distribution. More precisely than these simple remarks, it can be
proved that \emph{a L\'evy process $X(t)$ is selfsimilar if and only
if it is \SST}\ (see~\cite{sato} p.\ 71).

Things are rather different, however, when we consider only additive
(not necessarily L\'evy) processes, namely when we can also live
without stationarity. Now we must stick to the general
selfsimilarity equation\refeq{ssim}, and we must remark again that
there is another simple, intuitive procedure producing additive (but
not necessarily stationary), selfsimilar processes: simply consider
a \chf\ $\varphi(u)$, a real number $H>0$ and take
\begin{equation}\label{ssimchf}
    \phi_{s,t}(u)=\frac{\varphi\left(\left(t/T\right)^Hu\right)}
                       {\varphi\left(\left(s/T\right)^Hu\right)}\,.
\end{equation}
It is now apparent that the \chf s of this family satisfy the
equation\refeq{chk} and are also selfsimilar according to the
definition\refeq{ssim} with the space--time scale compensation
produced by $b=a^{H}$. Namely\refeq{ssimchf} produces an
$H$--selfsimilar process. Of course we must ask here the same
question surfaced w.r.t.\ equation\refeq{stchf} in the case of
stationary processes: when can we be sure that the function defined
by the ratio\refeq{ssimchf} of two \chf\ still is the bona fide
\chf\ of a law? Even in this case, however, the answer can be hinted
to by looking at the definition\refeq{sdlaw} of a \SD\ \chf$\,.$
More precisely it can be proved that\refeq{ssimchf} is a \chf\ if
and only if $\varphi(u)$ is \SD\ (see~\cite{sato} p.\ 99): this
ultimately brings out the intimate relation connecting
selfsimilarity and selfdecomposability.

Since \SD\ laws are also \ID\ the previous remarks show that from a
given \SD\ $\varphi(u)$ we can always produce two different kinds of
processes: a L\'evy process whose stationary increments follow the
law\refeq{stchf}; and a family of additive, selfsimilar process --
one for every value of $H>0$ -- whose (possibly non stationary)
increments follow the law\refeq{ssimchf}. All these processes
generated from the same $\varphi(u)$ are in general different with
one exception: when $\varphi(u)$ is an \aST\ law the associated
L\'evy process coincide with the $H$--selfsimilar one with Hurst
index $H=1/\alpha$. In this last case indeed the \chf s\refeq{stchf}
and\refeq{ssimchf} are identical (to see it take for example the
symmetric form\refeq{symmstable} of a \ST\ \chf). Remark also that
the index of an \aST\ law always satisfies $0<\alpha\leq2$
($\alpha=2$ for the Gaussian law), and that this is coherent with
the limitation $H=1/\alpha\geq1/2$ for the Hurst index of the
stable, selfsimilar processes (see~\cite{sato} p.\ 75). On the other
hand, in every other case (either non--\ST, or \aST\ with
$\alpha\neq1/H$), from a \SD\ law $\varphi(u)$ we can always build a
L\'evy, non selfsimilar process from\refeq{stchf}, and a family of
additive, selfsimilar processes with non stationary increments
from\refeq{ssimchf}. Finally from a \ID, but not \SD\ law we can
only get a L\'evy process from\refeq{stchf}, but no selfsimilarity
is allowed. For more details about present interest of the \SD\
distributions and selfsimilar processes in the applications see for
instance~\cite{carr} and~\cite{cufaro}.

Remark that selfsimilarity is not tied to the dependence or
independence of the increments: we have seen here that among
independent increment processes we find both selfsimilar and non
selfsimilar processes; and on the other hand a process can be
selfsimilar without showing independence of the increments. A
celebrated example of this second case is the so called
\emph{fractional Brownian motion}: this is a centered, Gaussian,
$H$--selfsimilar (for $H\in[0,1]$) process $B(t)$ with stationary
increments, and covariance function
\begin{equation*}
   \ave{B(t)B(s)}=\frac{|t|^{2H}+|s|^{2H}-|t-s|^{2H}}{2}\,,\qquad\quad t,s>0.
\end{equation*}
It is apparent that this is nothing else than a generalization of
the well known covariance function of the usual Brownian motion
$\ave{B(t)B(s)}=\min(t,s)$ that is recovered when $H=1/2$. Since
$B(t)$ is centered and Gaussian, this covariance function is all
that is needed to define the process also if it is not Markovian. In
fact a fractional Brownian motion coincides with the usual Brownian
motion (and hence is Markovian with independent increments) only for
$H=1/2$, while for $H\neq1/2$ it is non--Markovian, has correlated
increments and for $H>1/2$ shows long--range dependence. In other
words an $H$--selfsimilar fractional Brownian motion with $H\neq1/2$
is neither additive, nor Markovian: in fact it is not even a
semimartingale, and hence few results of stochastic calculus can be
used. From another standpoint (see~\cite{cont} p.\ 230) we can say
that the selfsimilarity can have different origins: it can stem
either from the length of the distribution tails of independent
increments, or from the correlation between short--tailed, Gaussian
increments, and the two effects can also be mixed. For more
information about the fractional Brownian motion see~\cite{mandel}
and~\cite{oks}

\subsection{Selfdecomposable laws and \ou\ processes}\label{ouproc}

Selfdecomposable laws appear also in another important context: they
are the most general class of stationary distributions of
\emph{processes of the Ornstein--Uhlenbeck type}. Take a L\'evy
process $Z(t)$ with generating triplet $(a^2,\mu,c)$ and \lch\
$\chi(u)$, and for $b>0$ consider the stochastic differential
equation (for simplicity we take $T=1$)
\begin{equation}\label{ousde}
    dX(t)=-bX(t)\,dt+dZ(t),\quad\qquad X(0)=X_0\quad\Pas
\end{equation}
whose exact meaning is rather in its integral form
\begin{equation*}
    X(t)=X_0-b\int_0^tX(s)\,ds+Z(t).
\end{equation*}
When $Z(t)$ is a Wiener process the equation\refeq{ousde} coincides
with the \sde\ of an ordinary, Gaussian \ou\ process; but the
equation\refeq{ousde} keeps the meaning of a well behaved \sde\ even
if $Z(t)$ is a generic, non Gaussian L\'evy process, and its
solution
\begin{equation*}
    X(t)=X_0e^{-bt}+\int_0^te^{b(s-t)}dZ(s).
\end{equation*}
will be called a process of the \ou\ type. Of course to give a rigorous sense to this solution we should define
our stochastic integrals for a generic L\'evy process $Z(t)$: since all L\'evy processes are semimartingales
(see~\cite{cont} p.\ 255), this can certainly be done, but we will skip this point referring the reader to the
existing literature (see~\cite{sato} and~\cite{cont}, or~\cite{protter} for an extensive treatment). We will
rather shift our attention to the possible existence of \emph{stationary distributions} for a process of the \ou\
type. In fact it is possible to show (see~\cite{sato} p.\ 108, and~\cite{cont} p.\ 485) that if
\begin{equation*}
    \int_{|x|\geq1}\log|x|\,\mu(dx)<+\infty
\end{equation*}
then the \ou\ process $X(t)$ solution of\refeq{ousde} has a
stationary distribution $\varphi(u)=e^{\psi(u)}$ which is \SD\ with
\lch
\begin{equation}\label{oulch1}
    \psi(u)=\int_0^{+\infty}\chi(ue^{-bt})\,dt
\end{equation}
and generating triplet $(\beta^2,\nu,\gamma)$ where
$\beta^2=a^2/2b$, $\gamma=c/b$, and -- according to the
equation\refeq{sdlevymeas} -- the absolutely continuous L\'evy
measure $\nu$ has a density
\begin{equation*}
    W(x)=\frac{k(x)}{|x|}=\frac{1}{b\,|x|}\times\left\{
                                     \begin{array}{ll}
                                       \mu\{[x,+\infty)\}, & \hbox{if $x>0$;} \\
                                       \mu\{(-\infty,x]\}, & \hbox{if $x<0$.}
                                     \end{array}
                                   \right.
\end{equation*}
Conversely for every \SD\ law $\varphi(u)$ there is a L\'evy process
$Z(t)$ such that $\varphi(u)$ is the stationary law of the \ou\
process driven by $Z(t)$. Remark also that by the simple change of
variable $s=ue^{-bt}$ the relation\refeq{oulch1} takes the form
\begin{equation}\label{oulch}
    \psi(u)=\frac{1}{b}\int_0^u\frac{\chi(s)}{s}\,ds\,,\qquad\quad\chi(u)=bu\psi'(u)
\end{equation}
which is well suited to the inverse problem of finding the L\'evy
noise of an \ou\ process for a prescribed \SD\ stationary
distribution.

\section{Examples}

\subsection{Families of laws}\label{exampleLaws}

We will consider now several families of distributions (for further
details see for example~\cite{cufaro} and references quoted
therein), all absolutely continuous, centered and symmetric, with a
space scale parameter $a>0$ which will of course span the types
since the centering parameters always vanish:
\begin{itemize}
    \item the type of the Normal laws $\mathcal{N}(a)$ with \pdf\ and \chf\
\begin{equation*}
    f(x)=\frac{e^{-x^2/2a^2}}{a\sqrt{2\pi}}\,,\quad\qquad\varphi(u)=e^{-a^2u^2/2}\,,
\end{equation*}
    and with variance $a^2$;
    \item the types (one for every $\lambda>0$) of the Variance--Gamma laws $\mathcal{VG}(\lambda,a)$ with
\begin{equation*}
    f(x)=\frac{(|x|/a)^{\lambda-\frac{1}{2}}K_{\lambda-\frac{1}{2}}(|x|/a)}{a2^{\lambda-1}\Gamma(\lambda)\sqrt{2\pi}}\,,
    \quad\qquad \varphi(u)=\left(\frac{1}{1+a^2u^2}\right)^\lambda,
\end{equation*}
    where $K_\nu(z)$ are the modified Bessel functions and
    $\Gamma(z)$ is the Euler Gamma function~\cite{abramowitz}; their
    variance $2\lambda a^2$ is always finite;
    \item the types (one for every $\lambda>0$) of the Student laws $\mathcal{T}(\lambda,a)$ with \pdf\ and \chf\
\begin{equation*}
    f(x)=\frac{1}{aB\left(\frac{1}{2},\frac{\lambda}{2}\right)}\,\left(\frac{a^2}{a^2+x^2}\right)^{\frac{\lambda+1}{2}},
    \quad\qquad
    \varphi(u)=\frac{2(a|u|)^{\lambda/2}K_{\lambda/2}(a|u|)}{2^{\lambda/2}\Gamma(\lambda/2)}\,,
\end{equation*}
    where $B(x,y)$ is the Euler Beta function~\cite{abramowitz}.
    Their variance is finite only for $\lambda>2$ and its value is
    $a^2/(\lambda-2)$.
\end{itemize}
Important particular types within the Variance--Gamma and the
Student families are respectively the Laplace (double exponential)
laws $\mathcal{L}(a)=\mathcal{VG}(1,a)$ with \pdf\ and \chf\
\begin{equation*}
    f(x)=\frac{e^{-|x|/a}}{2a}\,,\quad\qquad\varphi(u)=\frac{1}{1+a^2u^2}\,,
\end{equation*}
and finite variance $2a^2$, and the Cauchy laws
$\mathcal{C}(a)=\mathcal{T}(1,a)$ with
\begin{equation*}
    f(x)=\frac{1}{a\pi}\,\frac{a^2}{a^2+x^2}\,,\qquad\quad\varphi(u)=e^{-a|u|},
\end{equation*}
and divergent variance. Finally we will also consider in the
following another type of Student laws
$\mathcal{S}(a)=\mathcal{T}(3,a)$ with \pdf\ and \chf\
\begin{equation*}
    f(x)=\frac{2}{a\pi}\,\left(\frac{a^2}{a^2+x^2}\right)^2\,,\qquad\quad\varphi(u)=e^{-a|u|}(1+a|u|),
\end{equation*}
and finite variance $a^2$.

All the laws of our families are \SD\ (and hence \ID), but only
$\mathcal{N}(a)$ and $\mathcal{C}(a)$ are types of \aST\
distributions: more precisely $\mathcal{N}(a)$ laws are 2--\ST, and
$\mathcal{C}(a)$ are 1--\ST. On the other hand the family
$\mathcal{VG}(\lambda,a)$ is closed under convolution, while
$\mathcal{T}(\lambda,a)$ is not. Of course this does not mean that
the Variance-Gamma laws are \ST\ since $\mathcal{VG}(\lambda,a)$ is
not a unique type, and a convolution will mix different types with
different $\lambda$ values. The \ID\ laws $\mathcal{N}(a)$,
$\mathcal{L}(a)$, $\mathcal{C}(a)$ and $\mathcal{S}(a)$ are of
course entitled to their characteristic triplets
$(\beta^2,\nu,\gamma)$. Since they are all centered and symmetric we
have $\gamma=0$ for all of them. As for $\beta^2$ it can be seen
that for $\mathcal{L}(a)$, $\mathcal{C}(a)$ and $\mathcal{S}(a)$ we
have $\beta^2=0$ (in fact, in terms of the L\'evy--It\^o
decomposition, they generate so--called \emph{pure jump processes}),
while for $\mathcal{N}(a)$ we have $\beta^2=a^2$. As for the L\'evy
measures, on the other hand, we first of all have $\nu=0$ for the
laws $\mathcal{N}(a)$: from the point of view of the sample path
properties this simply means that -- at variance with the other
three cases under present investigation -- the L\'evy processes
generated by Gaussian distributions never make jumps. The L\'evy
measures of the other three cases are instead all absolutely
continuous and have the following densities $W(x)$
\begin{eqnarray*}
 &&   \frac{e^{-|x|/a}}{|x|}
 \qquad\qquad\qquad\qquad\qquad\qquad\qquad\qquad\qquad\qquad\qquad\:
                         \hbox{for $\mathcal{L}(a)$} \\
 &&   \frac{a}{\pi x^2}
 \qquad\qquad\qquad\qquad\qquad\qquad\qquad\qquad\qquad\qquad\qquad\quad\;\,
                         \hbox{for $\mathcal{C}(a)$} \\
 &&   \frac{a}{\pi x^2}\left[1-\frac{|x|}{a}
         \left(\sin\frac{|x|}{a}\>\mathrm{ci}\frac{|x|}{a}-\cos\frac{|x|}{a}\>\mathrm{si}\frac{|x|}{a}\right)\right]
           \qquad\qquad\qquad \hbox{for $\mathcal{S}(a)$}
\end{eqnarray*}
where the sine and the cosine integral functions for $x>0$
are~\cite{abramowitz}
\begin{equation*}
    \mathrm{si}\, x= -\int_x^{+\infty}\frac{\sin t}{t}\,dt\,,
    \qquad\quad\mathrm{ci}\, x= -\int_x^{+\infty}\frac{\cos t}{t}\,dt\,.
\end{equation*}
\begin{figure}
\begin{center}
\includegraphics*[width=9.0cm]{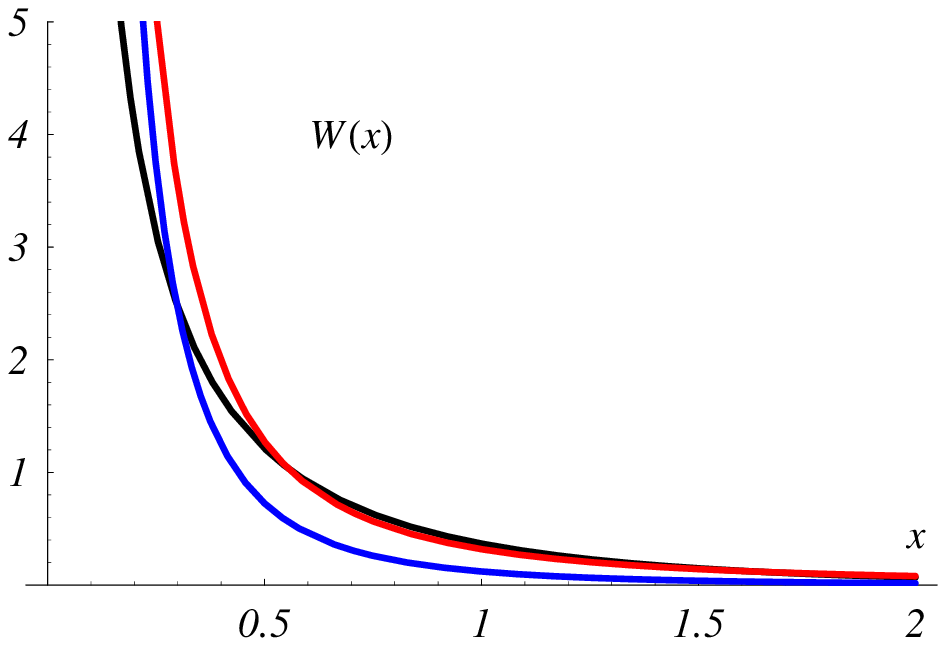}
\end{center}
\caption{\small Densities $W(x)$ of the L\'evy measures for the \SD\
laws $\mathcal{L}(a)$ (black), $\mathcal{C}(a)$ (red) and
$\mathcal{S}(a)$ (blue). To make the plots comparable we have chosen
$a=1$ for all the three densities, and to make the differences more
visible we have plotted only the positive $x$--axis since the curves
are exactly symmetric on the negative axis.}\label{fig01}
\end{figure}
Examples of these three densities are plotted in Figure~\ref{fig01}.

\subsection{Convergence of consecutive sums}\label{exampleCLT}

To give examples of consecutive sums converging to our laws let us first of all recall what happens in the case
of the Poisson laws $\mathcal{P}(\lambda)$. Let $\mathcal{B}(n,p)$ represent the binomial laws for $n$
independent trials of verification of an event occurring with probability $p$, and take the triangular array of
Bernoulli 0--1 \rv s $X_k^{(n)}\sim \mathcal{B}(1,\lambda/n)$ with $k=1,\ldots,n$ and $n=1,2,\ldots\,$ Apparently
they are \uan\ because for $0<\epsilon<1$ we have
\begin{equation*}
    \max_k\,\pr{|X_k^{(n)}|\geq\epsilon}=\frac{\lambda}{n}\ton0\,.
\end{equation*}
It is very well known that the consecutive sums are Binomial \rv s, namely
\begin{equation*}
    S_n=X_1^{(n)}+\ldots+X_n^{(n)}\sim \mathcal{B}\left(n,\frac{\lambda}{n}\right)
\end{equation*}
and that, according to the classical Poisson theorem, the limit law
of these $S_n$ is $\mathcal{P}(\lambda)$. Remark that, since in
passing from $n$ to $n'$ the \rv s $X_k^{(n)}$ change type, it will
not be possible to put $S_n$ in the form of a normed sum
as\refeq{normsum}. The Poisson laws, however, are also limit laws in
a still legitimate, but rather trivial sense due to the composition
properties of the family $\mathcal{P}(\lambda)$: take for instance a
triangular array of Poisson \rv s $X_k^{(n)}\sim
\mathcal{P}(\lambda/n)$ with $k=1,\ldots,n$ and $n=1,2,\ldots\,$
which are again \uan\ because for $0<\epsilon<1$ we have
\begin{equation*}
    \max_k\,\pr{|X_k^{(n)}|\geq\epsilon}=1-e^{-\lambda/n}\ton0\,.
\end{equation*}
Now -- at variance with the previous example of the Bernoulli
triangular array -- at every step $n$ we exactly have
$S_n\sim\mathcal{P}(\lambda)$ and hence, albeit in a trivial sense,
the limit law again is $\mathcal{P}(\lambda)$. Also in this case
$X_k^{(n)}$ and $X_k^{(n')}$ belong to different types, so that it
will be impossible to put $S_n$ in the form of a normed sum
as\refeq{normsum}. Of course both these examples show in what sense
the Poisson laws are \ID\ but not \SD: they are limit laws of
consecutive sums\refeq{consecsum} of \uan\ triangular arrays, but
not of normed sums\refeq{normsum} of independent \rv s.

At the other end of the gamut of the \ID\ laws we find the 2--\ST\
normal laws. To see in what sense they are limit laws take an
arbitrary sequence $X_k$ of centered, \iid\ \rv s with finite
variance $\sigma^2$ and define the triangular array
\begin{equation*}
    X_k^{(n)}=\frac{X_k}{\sigma\sqrt{n}}
\end{equation*}
which always turns out to be \uan\ because of the Chebyshev
inequality:
\begin{equation*}
    \max_k\,\pr{|X_k^{(n)}|\geq\epsilon}=\max_k\,\pr{|X_k|\geq\epsilon\sigma\sqrt{n}}
    \leq\frac{1}{\epsilon^2n}\ton0\,.
\end{equation*}
The consecutive sums are now also normed sums of \iid\ \rv s since
\begin{equation*}
    S_n=X_1^{(n)}+\ldots+X_n^{(n)}=\frac{1}{\sigma\sqrt{n}}\sum_{k=1}^nX_k
\end{equation*}
and according to the classical, normal Central Limit Theorem their
limit law is $\mathcal{N}(1)$. Also in this case, however, it is
possible to exploit the composition properties of the normal type to
find another, more trivial form of the consecutive sums: take the
sequence of normal \iid\ \rv s $X_k\sim\mathcal{N}(1)$ and define
the triangular array $X_k^{(n)}=X_k/\sqrt{n}\sim\mathcal{N}(1/n)$
which again apparently is \uan. Now the consecutive sums are also
normed sums which -- at variance with the previous example -- are
all normally distributed
\begin{equation*}
    S_n=X_1^{(n)}+\ldots+X_n^{(n)}=\frac{1}{\sqrt{n}}\sum_{k=1}^nX_k\sim\mathcal{N}(1)
\end{equation*}
and hence, in a trivial sense, the limit law is $\mathcal{N}(1)$.
Finally let us remark that, since the normal laws are \ID, nothing
will forbid them to be also limit laws of consecutive sums of
triangular arrays that \emph{do not} reduce to normed sums of \iid\
\rv s. However we will not elaborate here examples in this sense.

We have introduced the trivial forms of the consecutive sums in the
case of the Poisson and normal laws (see also the remarks at the end
of the Section~\ref{classicalth}) only because in our other
subsequent examples this will be the unique explicit form available
for our $S_n$. Remark also that these trivial forms essentially
derive from the fact that our laws are all \ID. In fact if $\varphi$
is \ID, then also $\varphi_n=\varphi^{1/n}$ is a \chf, and of course
$\varphi_n^n=\varphi$ for every $n$. In general -- with the
exception of the \ST\ laws -- the $\varphi_n$ are not of the same
type for different $n$, and hence the sums can not take the form of
normed sums. That notwithstanding, in a trivial sense, every \ID\
law $\varphi$ is the limit law of the consecutive sums of \iid\ \rv
s all distributed according to $\varphi_n$. What is less trivial,
however, is to give an explicit form to the \cdf\ or \pdf\ of the
component laws $\varphi_n$: as the subsequent examples will show
this can be easily done only when we deal with families of laws
closed under composition and decomposition.

Let us consider first the Cauchy laws introduced in the previous
Section~\ref{exampleLaws}: they are 1--\ST\ and, by taking advantage
of the fact that the family $\mathcal{C}(a)$ is closed under
composition and decomposition, it will be easy to show how they are
limit laws of suitable sums of \rv s. Take for instance a sequence
of \iid\ Cauchy \rv s $X_k\sim\mathcal{C}(a)$ and define the
triangular array $X_k^{(n)}=X_k/n\sim\mathcal{C}(a/n)$. Since now
there is no variance to speak about, to show that this sequence is
\uan\ we can not use the Chebyshev inequality. If however
\begin{equation*}
    F(x)= \frac{1}{2}+\frac{1}{\pi}\arctan\frac{x}{a}
\end{equation*}
is the common \cdf\ of the $X_k$, it is easy to see that the
sequence is \uan\ because
\begin{equation*}
    \max_k\,\pr{|X_k^{(n)}|\geq\epsilon}=\max_k\,\pr{|X_k|\geq\epsilon n}
    =2\left[1-F(\epsilon n)\right]\ton0\,.
\end{equation*}
Now, as for the Gaussian case, the consecutive sums are also normed
sums of \iid\ \rv s and are all distributed according to the Cauchy
law $\mathcal{C}(a)$
\begin{equation*}
    S_n=X_1^{(n)}+\ldots+X_n^{(n)}=\frac{1}{n}\sum_{k=1}^nX_k\sim\mathcal{C}(a)
\end{equation*}
so that, in a trivial sense, the limit law is $\mathcal{C}(a)$. What
forbids here the convergence to the normal law is the fact that the
variance is not finite, so that the normal Central Limit Theorem
does not apply. At variance with the Poisson and Gaussian previous
examples, however, we do not know non trivial forms of a Cauchy
limit theorem embodying the stability of the Cauchy law. In other
words we have neither explicit examples, nor general theorems
characterizing the form of the normed sums of \iid\ \rv s whose laws
converge to $\mathcal{C}(a)$, without being coincident with
$\mathcal{C}(a)$ at every step $n$ of the limiting process.

This last remark holds also in the case of the Laplace \SD, but not
\ST\ laws $\mathcal{L}(a)$. In fact, since the
$\mathcal{VG}(\lambda,a)$ family is closed under composition and
decomposition, we can always take a triangular array
$X_k^{(n)}\sim\mathcal{VG}(1/n,\,a)$ for $k=1,\ldots,n$ and
$n=1,2,\ldots$, and remark first that they are \uan\ by virtue of
the Chebyshev inequality (the $X_k^{(n)}$ have finite variance
$2a^2/n\ton0$), and then that for every $n$
\begin{equation*}
    S_n=X_1^{(n)}+\ldots+X_n^{(n)}\sim\mathcal{VG}(1,a)=\mathcal{L}(a)
\end{equation*}
so that the limit law trivially is $\mathcal{L}(a)$. It must also be
said that in this example the \rv s of the triangular array change
type with $n$ so that the corresponding consecutive sums $S_n$ can
not be recast in the form of normed sums of independent \rv s. Since
however the Laplace laws are not only \ID, but also \SD\ we would
expect to find normed sums of independent (albeit not \id, because
the Laplace laws are not \ST) \rv s whose laws converge to
$\mathcal{L}(a)$. Unfortunately we do not have general theorems
characterizing the needed sequences of independent \rv s, and we can
just show an example slightly more general than the previous one.
Take for instance the triangular array $X_k^{(n)}$ with laws
$\mathcal{VG}\left(\frac{1}{k(2+\log n)}\,,a\right)$ and variances
$\frac{a^2}{k(2+\log n)}$. They are \uan\ because from the Chebyshev
inequality we have
\begin{equation*}
    \max_k\,\pr{|X_k^{(n)}|\geq\epsilon}\leq\max_k\,\frac{a^2}{\epsilon^2k(2+\log n)}=\frac{a^2}{\epsilon^2(2+\log
    n)}\ton0\,,
\end{equation*}
while, for a known property of the harmonic numbers, the consecutive
sums are
\begin{equation*}
    S_n=X_1^{(n)}+\ldots+X_n^{(n)}\sim\mathcal{VG}\left(\frac{1}{2+\log
    n}\sum_{k=1}^n\frac{1}{k}\,,\,a\right)
    \ton\mathcal{L}(a)\,.
\end{equation*}
Now the sums $S_n$ are not trivially distributed according to
$\mathcal{L}(a)$ at every $n$, but again they can not be put in the
form of normed sums as they should since $\mathcal{L}(a)$ is \SD.

Finally similar remarks can be done for the \SD, but not \ST\
Student laws $\mathcal{S}(a)$ introduced in the
Section~\ref{exampleLaws}, but in this last case it is not even
possible to give a simple form to the trivial consecutive sums
because the Student family $T(\lambda,a)$ is not closed under
composition and decomposition. In other words if $\varphi$ is the
\chf\ of a law $\mathcal{S}(a)$ we are sure that
$\varphi_n=\varphi^{1/n}$ again is the \chf\ of a \ID\ law such that
$\varphi_n^n=\varphi$ for every $n$, but these component laws
$\varphi_n$ no longer belong to the $T(\lambda,a)$ family as happens
for the Variance--Gamma family, and in fact the form for their \pdf\
is rather complicated~\cite{cufaro}.

\subsection{Stationary and selfsimilar processes}\label{exampleproc}

Since all the laws of our examples are \ID\ we can use all of them
to generate the corresponding L\'evy processes by using\refeq{stchf}
to give the law of the increments on a time interval of width $t$.
In particular we will explicitly do that for the laws
$\mathcal{N}(a)$, $\mathcal{L}(a)$, $\mathcal{C}(a)$ and
$\mathcal{S}(a)$. We then get as stationary increment \chf s
$\phi_t(u)$ respectively
\begin{eqnarray*}
 && e^{-a^2tu^2/2T}   \qquad\qquad\qquad\qquad\;\;\, \hbox{\textit{Wiener process} from $\mathcal{N}(a)$}\\
 && \left(1+a^2u^2\right)^{-t/T}   \qquad\qquad\qquad\;\, \hbox{\textit{Laplace process} from $\mathcal{L}(a)$} \\
 && e^{-at|u|/T}   \qquad\qquad\qquad\qquad\quad\;\: \hbox{\textit{Cauchy process} from $\mathcal{C}(a)$} \\
 && e^{-at|u|/T}(1+a|u|)^{t/T}   \qquad\qquad \hbox{\textit{Student process} from $\mathcal{S}(a)$}
\end{eqnarray*}
It is then apparent that the laws of the increments for the
$\alpha$--\ST\ Wiener and Cauchy processes are simply
$\mathcal{N}(a\sqrt{t/T})$ and $\mathcal{C}(at/T)$, while for the
Laplace process the law of the increments is actually a Laplace law
only for $t=T$, while in general it is a $\mathcal{VG}(t/T,a)$ at
other values of $t$. For our Student process, on the other hand, the
situation is less simple because the Student family is not closed
under convolution, and the increment law no longer is in
$T(\lambda,a)$ for $t\neq T$. In this case it is not easy to find
the actual distribution from its Fourier transform $\phi_t(u)$, and
only recently it has been suggested that the increments are
distributed according to a mixture of other Student laws (for
further details see~\cite{cufaro}). The Wiener and the Cauchy
processes are $H$--selfsimilar with $H=1/2$ and $H=1$ respectively.
This can also be seen by looking at the interplay between the two --
spatial and temporal -- scale parameters $a$ and $T$. In fact in the
Wiener and Cauchy processes these two scale parameters appear in two
combinations -- respectively $a^2/T$ and $a/T$ -- such that a change
in the time units can always be compensated by a corresponding,
suitable change in the space units; as a consequence the
distribution of the process is left unchanged by these twin scale
changes. This, on the other hand, would not be possible in the
Laplace and Student processes since $a$ and $T$ no longer appear in
such combinations.

That notwithstanding we can achieve selfsimilarity in additive, non
stationary processes produced by all our \SD\ laws.
From\refeq{ssimchf} in fact we can give the \chf\ $\phi_{s,t}(u)$ of
the increments in the interval $[s,t]$ for our four types of law.
First of all from the Normal type $\mathcal{N}(a)$ we get
\begin{equation*}
    \phi_{s,t}(u)=e^{-a^2\left(t^{2H}-s^{2H}\right)u^2/2T^{2H}}
\end{equation*}
namely
\begin{equation*}
    X(t)-X(s)\sim\mathcal{N}\left(a^2\,\frac{t^{2H}-s^{2H}}{T^{2H}}\right)\,.
\end{equation*}
These laws define processes which coincide with the usual Wiener
process if and only if $H=1/2$. For $H\neq1/2$, on the other hand,
our process is additive, $H$--selfsimilar, and Gaussian with non
stationary increments, and hence does not even coincide with a
fractional Brownian motion which has stationary and correlated
increments. In a similar way from the laws of the Cauchy type
$\mathcal{C}(a)$ we get
\begin{equation*}
    X(t)-X(s)\sim\mathcal{C}\left(a\,\frac{t^{H}-s^{H}}{T^{H}}\right)\,,
\end{equation*}
and the process will coincide with the stationary (L\'evy) Cauchy
process when $H=1$, while for $H\neq1$ we have an additive,
$H$--selfsimilar process with non stationary increments. From the
Laplace type $\mathcal{L}(a)$ on the other hand we obtain an
$H$--selfsimilar (with $H>0$), additive process when we take
\begin{equation*}
    \phi_{s,t}(u)=\frac{1+a^2\left(\frac{s}{T}\right)^{2H}u^2}
                         {1+a^2\left(\frac{t}{T}\right)^{2H}u^2}\:=\:
                         \left(\frac{s}{t}\right)^{2H}+\left[1-\left(\frac{s}{t}\right)^{2H}\right]
                         \frac{1}{1+a^2\left(\frac{t}{T}\right)^{2H}u^2}\,;
\end{equation*}
so that, for $s>0$, the law of the increment on an interval $[s,t]$
actually is a mixture -- with time--dependent probabilistic weights
-- of a law degenerate in $x=0$ and of a Laplace law:
\begin{equation*}
    X(t)-X(s)\sim\left(\frac{s}{t}\right)^{2H}\delta_0+\left[1-\left(\frac{s}{t}\right)^{2H}\right]
                               \mathcal{L}\left(\frac{a^2t^{2H}}{T^{2H}}\right)\,.
\end{equation*}
In other words this means that for $s>0$ there is always a non--zero
probability that in $[s,t]$ the process increment will simply
vanish. Finally from the Student type $\mathcal{S}(a)$ we get the
additive, $H$--selfsimilar, non stationary process with
\begin{equation*}
    \phi_{s,t}(u)=\frac{e^{-at^{H}|u|/T^{H}}\left(1+\frac{at^H}{T^H}|u|\right)}
                       {e^{-as^{H}|u|/T^{H}}\left(1+\frac{as^H}{T^H}|u|\right)}\,,
\end{equation*}
so that $X(t)\sim\mathcal{S}\left(at^H/T^H\right)$, while nothing
simple enough can be said of the independent increment laws.

\subsection{\ou\ stationary distributions}

All the L\'evy processes introduced in the Section~\ref{exampleproc}
can now be used as driving noises of \ou\ processes according to the
discussion of the Section~\ref{ouproc}. Here we will only list the
essential properties of the corresponding stationary distributions
by analyzing their \lch s\refeq{oulch}. First of all, if $b$ is the
parameter of the process as in\refeq{ousde} (remember that we took
there $T=1$ for simplicity), for the Wiener and the Cauchy driving
noises we immediately have from\refeq{oulch} that the \lch s
$\psi(u)$ of the stationary distributions are respectively
\begin{eqnarray*}
   && -\frac{a^2u^2}{4b}\,,\qquad\qquad\quad \hbox{for a Wiener noise (usual \ou\ process)}\\
   && -\frac{a|u|}{b}\,,\qquad\qquad\quad\; \hbox{for a Cauchy noise}
\end{eqnarray*}
namely that the stationary distributions simply are
$\mathcal{N}(a/\sqrt{2b})$ and $\mathcal{C}(a/b)$. In the case of
the Wiener noise (namely in the case of the ordinary, Gaussian \ou\
process) this means that the stationary distribution has a variance
$a^2/2b$ while the Gaussian law generating the Wiener process had a
variance $a^2$. For the Cauchy noise, on the other hand, there is no
variance to speak about.

For the other two \ou\ L\'evy noises (Laplace and Student) an
important role is played by the so called \emph{dilogarithm}
function~\cite{abramowitz,lewin,morris}
\begin{equation*}
    \mathrm{Li}_2(x)=\int_x^0\frac{\log(1-s)}{s}\,ds\,,
    \qquad\left(=\sum_{k=1}^\infty\frac{x^k}{k^2}\,,\qquad|x|\leq1\right)
\end{equation*}
In fact a direct calculation of the integrals\refeq{oulch} gives for
the \lch s $\psi(u)$
\begin{eqnarray*}
   && \frac{1}{2b}\,\mathrm{Li}_2(-a^2u^2)\,,\qquad\qquad\qquad\qquad\; \hbox{for a Laplace noise}\\
   && -\frac{a|u|}{b}-\frac{1}{b}\,\mathrm{Li}_2(-a|u|)\,,\qquad\qquad\quad\; \hbox{for a Student noise}
\end{eqnarray*}
From the \chf s $\varphi(u)=e^{\psi(u)}$ we can also calculate the
stationary variances as $-\varphi''(0)$ and we get $a^2/b$ and
$a^2/2b$ respectively in the Laplace and in the Student case. Remark
that -- when they exist finite -- the variances of the stationary
distributions always are in the same relation with the variance of
the law generating the noise: the stationary variance is the
generating law variance divided by $2b$. The form of the
corresponding \pdf s is not known analytically, but it can be
assessed by numerically calculating the inverse Fourier transforms
of the \chf s: the results of these calculations for a couple of
particular cases are shown in the Figures~\ref{fig04}
and~\ref{fig05}.

Finally, since the types of laws analyzed in this section are all
\SD, by reversing the previous procedure we can also add a few
remarks about the \ou\ driving noises required to have
$\mathcal{N}(a)$, $\mathcal{L}(a)$, $\mathcal{C}(a)$ and
$\mathcal{S}(a)$ as stationary distributions. In fact we have
already said that for our two $\alpha$--stable cases the stationary
distributions are of the same type of the laws generating the
driving noise, so that there is essentially nothing to add for the
$\mathcal{N}(a)$ and $\mathcal{C}(a)$ stationary distributions. As
for the other two cases on the other hand we will use the second
equation\refeq{oulch} to calculate the noise \lch s $\chi(u)$:
\begin{figure}
\begin{center}
\includegraphics*[width=13.0cm]{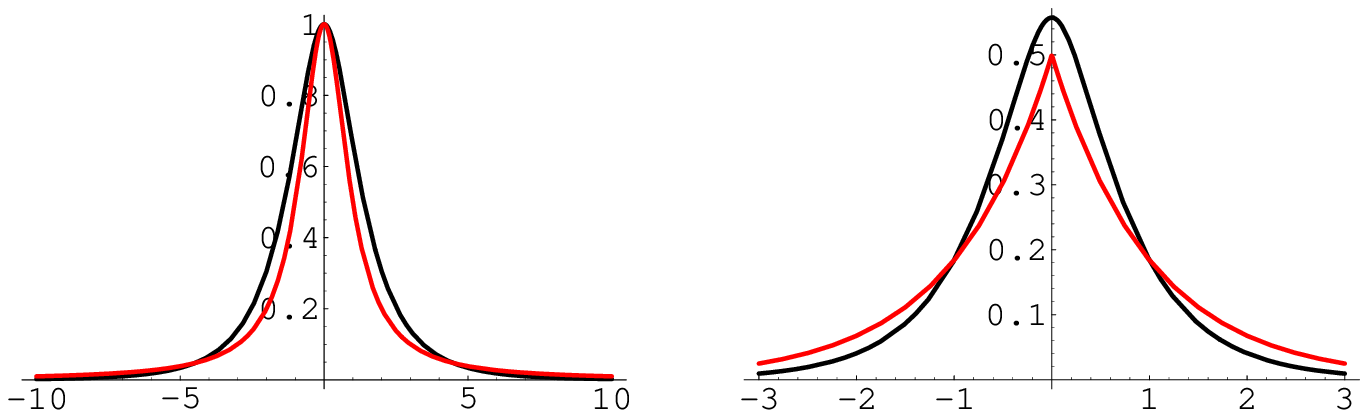}
\end{center}
\caption{\small \ou\ process driven by a Laplace noise: \chf s
(left) and \pdf s (right) of the stationary distribution (black
lines), compared with the \chf s and \pdf s of the Laplace law
$\mathcal{L}(a)$ generating the driving noise (red lines). Here
$a=1/\sqrt{2}$ and $b=1$ so that both the variances (that of the
stationary distribution, and that of $\mathcal{L}(a)$) are equal to
1.}\label{fig04}
\begin{center}
\includegraphics*[width=13.0cm]{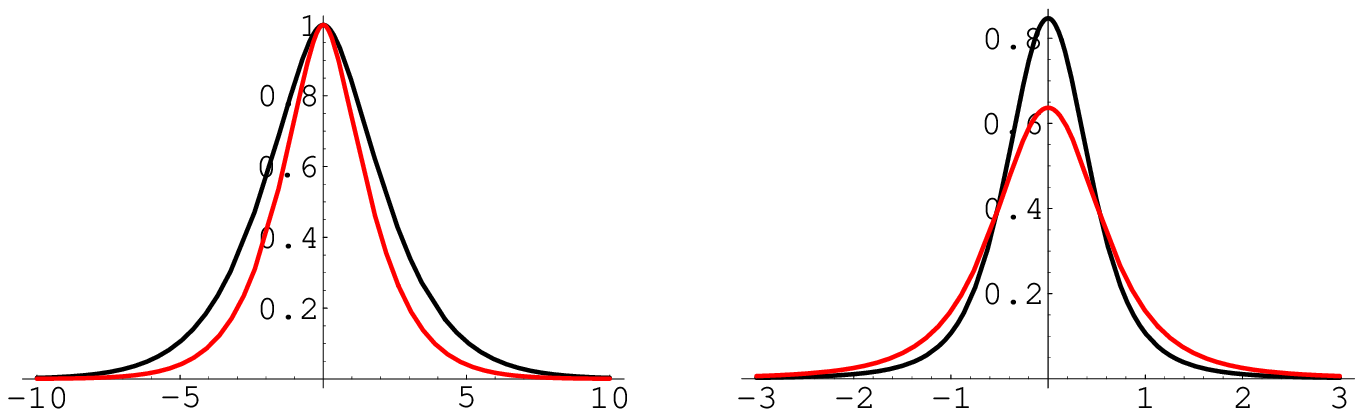}
\end{center}
\caption{\small \ou\ process driven by a Student noise: \chf s
(left) and \pdf s (right) of the stationary distribution (black
lines), compared with the \chf s and \pdf s of the Student law
$\mathcal{S}(a)$ generating the driving noise (red lines). Here
$a=b=1$ so that both the variances (that of the stationary
distribution, and that of $\mathcal{S}(a)$) are equal to
1.}\label{fig05}
\end{figure}
\begin{eqnarray*}
   && -2b\,\frac{a^2u^2}{1+a^2u^2}\,,\qquad\qquad\qquad\qquad\:
               \hbox{for a Laplace $\mathcal{L}(a)$ stationary law}\\
   && -b\,\frac{a^2u^2}{1+a|u|}\,,\qquad\qquad\qquad\qquad\quad
               \hbox{for a Student $\mathcal{S}(a)$ stationary law}
\end{eqnarray*}
In both cases however the \chf s can not be elementarily inverted so
that we do not have an explicit expression for the increment \pdf s
of the the L\'evy noises that produce these two stationary \ou\
distributions. We can only add a few remarks about the Laplace case:
here the \chf\ $\varphi(u)=e^{\psi(u)}$ does not vanish at the
infinity since $\varphi(\pm\infty)=e^{-2b}$. As a consequence the
L\'evy noise increment \chf\ can be better written as
\begin{equation*}
    [\varphi(u)]^t=e^{-2bta^2u^2/(1+a^2u^2)}=e^{-2bt}+(1-e^{-2bt})\,\frac{e^{2bt/(1+a^2u^2)}-1}{e^{2bt}-1}
\end{equation*}
so that the independent increments of an \ou\ process with the
Laplace type $\mathcal{L}(a)$ as stationary law are distributed
according to a time--dependent mixture of two laws, one of which is
degenerate in $x=0$. As for the second law of this mixture, it has a
\pdf\ given by
\begin{equation*}
    f(x,t)=\frac{1}{\pi}\int_0^{+\infty}\cos(ux)\,\frac{e^{2bt/(1+a^2u^2)}-1}{e^{2bt}-1}\,\,du
\end{equation*}
but this integration can not be analytically performed.

\section{Conclusions}

Since many years selfsimilarity is a fashionable subject of
investigation, in areas ranging from fractals to long--range
interactions in complex systems: to have an idea just ask for the
papers with the word ``self similarity'' either in their title or in
their abstract present on \verb"arxiv.org" and you will find
$1\,000$ articles, and almost 200 of them only in the first six
months of 2007. On the other hand this is a subject that has been
approached from many different standpoints producing an unavoidable
level of confusion~\cite{mccauley}, while in fact it would be better
discussed by placing the reader in the perspective of the general
theory of the infinitely divisible (even non stable) processes. In
the field of mathematical finance the use of non stable L\'evy
processes is widespread, and several families of selfdecomposable
laws and processes have been intensively studied in recent years:
see for example the case of the Generalized Hyperbolic
family~\cite{eberlein1,raible,eberlein2}, of the Student
family~\cite{cufaro,heyde}, and of the Variance Gamma
family~\cite{madan1,madan2,madan3,madan4}. Considerable interest has
also been elicited by the use of selfdecomposable laws in connection
with the \ou\ processes~\cite{cont,carr}, in particular for the
stochastic volatility modelling. However, while in econophysics some
non stable L\'evy laws are recognized as possible candidates for a
consistent modelling of the underlying processes~\cite{bouchaud,
mantegna}, they remain less popular in the field of statistical
mechanics and only recently their use has been proposed in
connection with applications to the technology of accelerator
beams~\cite{cufaro,cufaro1,cufaro2}

In this paper we have tried to elucidate just a few points in the
framework of the theory of stochastic processes. In particular we
focused our attention on the relation between on the one hand the
selfsimilarity, and on the other the independence and the
stationarity of the increments. This has led our inquiry toward the
analysis of the laws of the process increments, and we have stressed
the connection between the selfsimilarity of the process and the
selfdecomposability of the increment laws. Selfdecomposable laws
naturally arise in the study of the Central Limit Problem and of its
solutions: in fact they are an intermediate (and more elusive) class
of distributions between the more general infinitely divisible, and
the more particular (and more popular) stable distributions. We
found then that, in the case of selfdecomposable generating laws,
both the stationarity of the increments and the selfsimilarity are
always possible, but are not always present in the same process. On
the other hand selfsimilarity can also be a property of (non
Markovian) processes with non independent increments as in the case
of the fractional Brownian motion. We finally stressed the
connection between the selfdecomposability and the stationary laws
of generalized \ou\ processes with non Gaussian, L\'evy noises. All
that has also be elucidated by means of a few particular examples,
and some kind of application from physics to finance has also been
pointed out.

\vfill\eject

\end{document}